\documentclass[manuscript]{aastex}

\usepackage{epsfig}                   
\usepackage{multirow}

\shorttitle{$^{25}$Mg(p,$\gamma$)$^{26}$Al reaction}
\shortauthors{LUNA}

\begin{document}
\title{Impact of a revised $^{25}$Mg(p,$\gamma$)$^{26}$Al reaction rate on the operation of the Mg-Al cycle}

\author{O.~Straniero}
     \affil{INAF-Osservatorio Astronomico di Collurania, Teramo, Italy,
      and INFN Sezione di Napoli, Napoli, Italy}
      \email{straniero@oa-teramo.inaf.it}
\author{G.~Imbriani}
    \affil{Dipartimento di Scienze Fisiche, Universit\`a di Napoli ''Federico II'', and INFN Sezione di Napoli, Napoli, Italy}
\author{F.~Strieder}
    \affil{Institut f\"ur Experimentalphysik, Ruhr-Universit\"at Bochum, Bochum, Germany}
\author{D.~Bemmerer}
    \affil{Helmholtz-Zentrum Dresden-Rossendorf, Bautzner Landstr 400, Germany}
\author{C.~Broggini}
     \affil{Istituto Nazionale di Fisica Nucleare (INFN), Sezione di Padova, via Marzolo 8, 35131 Padova, Italy}
\author{A.~Caciolli}
     \affil{Istituto Nazionale di Fisica Nucleare (INFN), Sezione di Padova, via Marzolo 8, 35131 Padova, Italy and Legnaro National Laboratory (INFN), Legnaro (Padova), Italy}
\author{P.~Corvisiero}
     \affil{Universit\`a di Genova and INFN Sezione di Genova, Genova, Italy}
\author{H.~Costantini}
     \affil{Universit\`a di Genova and INFN Sezione di Genova, Genova, Italy}
\author{S. Cristallo}
     \affil{INAF-Osservatorio Astronomico di Collurania, Teramo, Italy,
      and INFN Sezione di Napoli, Napoli, Italy}
\author{A.~DiLeva}
     \affil{Dipartimento di Scienze Fisiche, Universit\`a di Napoli ''Federico II'', and INFN Sezione di Napoli, Napoli, Italy}
\author{A.~Formicola}
    \affil{INFN, Laboratori Nazionali del Gran Sasso (LNGS), Assergi (AQ), Italy}
\author{Z.~Elekes}
     \affil{Institute of Nuclear Research (ATOMKI), Debrecen, Hungary}
\author{Zs.~F\"ul\"op}
     \affil{Institute of Nuclear Research (ATOMKI), Debrecen, Hungary}
\author{G.~Gervino}
     \affil{Dipartimento di Fisica Universit\`a di Torino and INFN Sezione di Torino, Torino, Italy}
\author{A.~Guglielmetti}
     \affil{Universit\`a degli Studi di Milano and INFN, Sezione di Milano, Italy}
\author{C.~Gustavino}
     \affil{INFN, Laboratori Nazionali del Gran Sasso (LNGS), Assergi (AQ), Italy}
\author{Gy.~Gy\"urky}
     \affil{Institute of Nuclear Research (ATOMKI), Debrecen, Hungary}
\author{M.~Junker}
    \affil{INFN, Laboratori Nazionali del Gran Sasso (LNGS), Assergi (AQ), Italy}
\author{A.~Lemut}\altaffilmark{1}
     \affil{Universit\`a di Genova and INFN Sezione di Genova, Genova, Italy}
\author{B.~Limata}
    \affil{Dipartimento di Scienze Fisiche, Universit\`a di Napoli ''Federico II'', and INFN Sezione di Napoli, Napoli, Italy}
\author{M.~Marta}\altaffilmark{2}
     \affil{Helmholtz-Zentrum Dresden-Rossendorf, Bautzner Landstr 400, Germany}
\author{C.~Mazzocchi}
     \affil{Universit\`a degli Studi di Milano and INFN, Sezione di Milano, Italy}
\author{R.~Menegazzo}
     \affil{Istituto Nazionale di Fisica Nucleare (INFN), Sezione di Padova, via Marzolo 8, 35131 Padova, Italy}
\author{L. Piersanti}
     \affil{INAF-Osservatorio Astronomico di Collurania, Teramo, Italy,
      and INFN Sezione di Napoli, Napoli, Italy}
\author{P.~Prati}
     \affil{Universit\`a di Genova and INFN Sezione di Genova, Genova, Italy}
\author{V.~Roca}
     \affil{Dipartimento di Scienze Fisiche, Universit\`a di Napoli ''Federico II'', and INFN Sezione di Napoli, Napoli, Italy}
\author{C.~Rolfs}
     \affil{Institut f\"ur Experimentalphysik, Ruhr-Universit\"at Bochum, Bochum, Germany}
\author{C.~Rossi Alvarez}
     \affil{Istituto Nazionale di Fisica Nucleare (INFN), Sezione di Padova, via Marzolo 8, 35131 Padova, Italy}
\author{E.~Somorjai}
     \affil{Institute of Nuclear Research (ATOMKI), Debrecen, Hungary}
\author{F.~Terrasi}
     \affil{Seconda Universit\`a di Napoli, Caserta, and INFN Sezione di Napoli, Napoli, Italy}
\author{H.-P.~Trautvetter}
    \affil{Institut f\"ur Experimentalphysik, Ruhr-Universit\"at Bochum, Bochum, Germany}

\altaffiltext{1}{present address: Lawrence Berkeley National Laboratory, Berkley, CA 94720 USA}
\altaffiltext{2}{present address: GSI Helmholtzzentrum f\"ur Schwerionenforschung GmbH, 64291 Darmstadt, Germany}

\date{\today}

\begin{abstract}
Proton captures on Mg isotopes play an important role in the Mg-Al cycle
active in stellar H-burning regions.
In particular, low-energy nuclear resonances
in the $^{25}$Mg(p,$\gamma$)$^{26}$Al reaction affect
the production of radioactive $^{26}$Al$^{gs}$ as well as the resulting
Mg/Al abundance ratio.
Reliable estimations of these quantities require precise measurements
of the strengths of low-energy resonances. Based on a new experimental
study performed at LUNA, we provide revised rates of the
$^{25}$Mg(p,$\gamma$)$^{26}$Al$^{gs}$
and the $^{25}$Mg(p,$\gamma$)$^{26}$Al$^{m}$ reactions with corresponding uncertainties.
In the temperature range 50 to
150 MK, the new recommended rate of the $^{26}$Al$^{m}$
production is up to 5 times higher than previously assumed.
In addition, at T$=100$ MK, the revised total reaction rate is a
factor of 2 higher. Note that this is the range of temperature at which the Mg-Al 
cycle operates in an H-burning zone. 

The effects of this revision are discussed. 
Due to the significantly larger
$^{25}$Mg(p,$\gamma$)$^{26}$Al$^{m}$
rate, the estimated production of $^{26}$Al$^{gs}$ in H-burning regions
is less efficient than previously obtained. As a result,
the new rates should imply a smaller contribution from Wolf-Rayet stars
to the galactic $^{26}$Al budget. Similarly, we show that
the AGB extra-mixing scenario does not appear able to explain
the most extreme values of $^{26}$Al/$^{27}$Al, i.e. $>10^{-2}$, found in some
O-rich presolar grains.
Finally, the substantial increase of the total reaction rate makes 
the hypothesis of a self-pollution by massive AGBs a more
robust explanation for the Mg-Al anticorrelation observed in 
Globular-Cluster stars.
\end{abstract}

\keywords{Nuclear reactions, nucleosynthesis, abundances --- Stars: AGB and post-AGB ---
Stars: Wolf-Rayet --- globular clusters: general}
\maketitle

\section{Introduction}

Many important astronomical phenomena are related to the occurrence of
the Mg-Al cycle in stellar interiors.
In past decades several potential stellar sites with an active
Mg-Al cycle have been identified. In particular, this cycle is active
in the deepest layer of a H-burning zone provided
the temperature is sufficiently large ($\rm T>40$ MK). Therefore, the necessary conditions
are fulfilled in the core of massive main sequence stars ($\rm M>30$~M$_\odot$)
as well as in the H-burning shells of red giant branch (RGB),
asymptotic giant branch (AGB) and red super-giant stars.
The Mg-Al cycle is also active during explosive H-burning events,
such as Nova like outbursts.

In these stars the H burning is often coupled to extended
mixing episodes, such as mixing powered by convection or other physical
processes, e.g. rotational induced instabilities, so that 
the products of the internal nucleosynthesis may appear 
at the stellar surface and can be directly observed.
In addition, these stars undergo intense mass loss episodes and, thus, they provide
an important contribution to the pollution of the interstellar medium.   
The presence of radioactive $^{26}$Al
(ground state half-live $t_{1/2}\approx7\times10^5$ yr)
in different astronomical environments may be a trace 
of the operation of the Mg-Al cycle in
stellar interiors. For example, the detection of the 1.809 MeV $\gamma$-ray line demonstrates
that a few M$_\odot$ of this isotope is presently alive in the galactic disk
\citep[see][]{diehl2006}. On the other hand,
the excess of $^{26}$Mg in the solar system material,
proves that some
radioactive $^{26}$Al has been injected into the presolar nebula shortly
before the solar system formation, about 4.5 Gyr ago \citep{lee77,gallino04}.
Furthermore, a $^{26}$Mg
excess has also been found in several presolar grains, such as
SiC grains belonging to the so-called mainstream type \citep{zinner1991}.
These grains most likely condensed in atmospheres of C-rich AGB stars and,
therefore, are believed to be fingerprints of the chemical composition
of these stars.
Finally, an evidence of the operation of the Mg-Al cycle is commonly found
in Globular Cluster stars, which show a clear anticorrelation between
Mg and Al \citep{kraft,gratton2001}.
This anticorrelation is usually ascribed to an early pollution
(occurred about 13 Gyr ago) of the
intra-cluster gas caused by massive AGB, perhaps super-AGB stars \footnote{In the following, 
with massive AGB we refer to stars with initial mass between $\sim5$ and $\sim8$ M$_\odot$. 
After the core-He burning, they form a degenerate C-O core and experience an AGB phase. 
With super-AGB we 
refer to stars with initial mass between $\sim8$ and $\sim10$ M$_\odot$. These stars ignite 
carbon in the degenerate core (usually it is an off-center ignition,
due to the plasma neutrino cooling), form an O-Ne
core and enter the super-AGB phase \citep{ritossa}. Note that the exact values of these 
mass limits depend on the chemical composition and their theoretical derivation is 
significantly affected by the uncertainties of several inputs physics.}

An accurate understanding of the stellar sites where the Mg-Al
cycle takes place may provide solutions for many open issues in stellar
evolution, stellar nucleosynthesis as well as chemical evolution.
Spectroscopic observations of Al and Mg coupled to information of the $^{26}$Al
radioactive decay derived from direct observations, e.g. $\gamma$-ray astronomy,
or indirect measures, e.g. isotopic analysis of solar system and presolar
material, may constrain stellar models in a wide range of stellar masses
and evolutionary phases. These correlations provide unique opportunities to
study the coupling between mixing processes and nuclear burning.

However, this work requires a precise evaluation of the nuclear reaction
rates of the Mg-Al cycle.
As part of a long-lasting experimental campaign on H-burning reactions,
the LUNA collaboration has recently measured the 
$^{25}$Mg(p,$\gamma)^{26}$Al rate at the Gran Sasso National Laboratory \citep{Strieder12_PLB}.
In the present work we use this new mesurements to revise the rate of
this important Mg-Al cycle reaction. In the next section we briefly
review the status of the experimental data in the relevant astrophysical
energy region and recommend a set of nuclear physics parameters that should
be used for the reaction rate calculations. As shown in Figure \ref{mgalcycle}, 
the proton capture on $^{25}$Mg may
produce $^{26}$Al in two different states, namely the ground state and
the isomeric state at ${\rm E_x}=228$ keV.
The corresponding reaction rates are provided in 
 section 3 as a function of temperature.
A final discussion follows, where some of the possible
astrophysical applications are addressed.

\section{Experimental Studies of the reaction $^{25}$M\lowercase{g(p},$\gamma$)$^{26}$A\lowercase{l}}

The astrophysical reaction rate of $^{25}$Mg(p,$\gamma$)$^{26}$Al (Q = 6.306~MeV) is dominated by narrow resonances. These resonances have been studied in previous experiments down to a low-energy limit of $\rm E = 189$~keV \citep{CH83a,CH83b,CH86,CH89,EN86,EN88,EN87,IL90,EN90,IL96,PO98}. The known $^{26}$Al level structure suggested the existence of additional low-lying resonances at $\rm E = 37$, 57, 92, 108, and 130~keV, among which the 92 keV resonance appears most important for astrophysical temperatures from 50 to 120 MK. 
These low-energy resonances, indeed, were identified in indirect experiments through transfer reaction studies \citep[see][and references therein]{IL96}.

Recently, in an experiment at the underground 400~kV LUNA (Laboratory for Underground Nuclear Astrophysics) accelerator in the Laboratori Nazionali del Gran Sasso \citep{review,Broggini10_ARNPS} the resonance at 92~keV was for the first time observed in a direct study \citep{Strieder12_PLB}. The resonance strengths of the 92, 189, and 304~keV resonances have been measured with unprecedented sensitivity taking full advantage of the extremely low $\gamma$-ray background level in the Gran Sasso laboratory. The Gran Sasso underground laboratory, where an average rock coverage of 1400 m (3800 meter water equivalent) reduces the $\gamma$-ray background signal by several orders of magnitude \citep{review}, is the ideal location for measurements of many astrophysically important nuclear reactions. In spite of tremendous experimental efforts in background reduction, target sample preparation as well as improvements in $\gamma$-ray detection, other low-energy resonances are still unaccessible for direct detection.

\subsection{The resonance strengths}

The strength of a resonance is defined in terms of nuclear resonance parameters:

\begin{equation}
\omega\gamma=\frac{2J+1}{(2j_1+1)(2j_2+1)}\frac{\Gamma_a\Gamma_b}{\Gamma}
\end{equation}
with $J$ , $j_1$, $j_2$ the spins of resonance, projectile and target nucleus, respectively, and  $\Gamma_a$, $\Gamma_b$, $\Gamma$
the partial widths for the entrance and exit channel, and the total resonance width at the resonance energy, respectively.
The resonance strength for narrow resonances as in the case of $^{25}$Mg(p,$\gamma$)$^{26}$Al can be measured directly in the thick-target yield approximation \citep[see][for details]{Rolfs88}. Alternatively, the resonance parameters, e.g. the proton partial width $\Gamma_p$ of the entrance channel, can be obtained from indirect experiments (see below).


The determination of weak low-energy resonance strengths from direct measurements is usually extremely difficult. Small target contaminations, e.g. oxygen, as well as stoichiometry changes under heavy proton bombardment may have a large effect on the absolute determination. A measurement relative to a well-known resonance can often avoid such difficulties. In \citet{Strieder12_PLB} the low-energy resonances have been normalized to the 304 keV resonance which in turn was precisely measured with several different experimental techniques \citep{Limata10_PRC}. The resonance strength values used for the present reaction rate calculation are summarized in Table \ref{omegagamma_comparison} and compared to NACRE \citep{angulo} and a more recent compilation by \citet{Iliadis10_NPA2}. Additionally, the ground state feeding probability and the electron screening correction for directly measured $\omega\gamma$ values are listed.

\subsection{Indirect experiments}

The NACRE \citep{angulo} rate at low temperatures (resonances below 130~keV) is mainly based on a reanalysis \citep{IL96} of proton partial width values from older proton stripping data \citep{Betts78_NPA,CH89,Rollefson90_NPA}. The same source of information was used in \citet{Iliadis10_NPA2}.

The proton width of the 92~keV resonance calculated from the recent direct experiment, $\Gamma_{\rm p}=(5.6\pm1.1)\times10^{-10}$~eV \citep{Strieder12_PLB}, deviates from the value used in the compilations, $\Gamma_{\rm p}=(2.8\pm1.1)\times10^{-10}$~eV \citep{Iliadis10_NPA3}, by 1.8$\sigma$. Therefore, at the 90~\% confidence level the two values are incompatible, while the proton width of \citet{Strieder12_PLB} is in good agreement with the original value of \citet{Rollefson90_NPA}, $\Gamma_{\rm p}=(5.2\pm1.3)\times10^{-10}$~eV. In contrast to the 92~keV resonance where a large spread of the indirect data is obvious (see Table II in \citet{IL96}), the proton width data for the 37 and 57~keV resonances from different experiments are in much better agreement and we used the value quoted in \citet{Iliadis10_NPA3} for the present work. As a general rule we have used data from direct experiments whenever available and the results from indirect measurements were included only where no direct data exists. 

\subsection{The ground state feeding factor}\label{gsfeeding}

The $^{25}$Mg(p,$\gamma$)$^{26}$Al resonances decay through complex $\gamma$-ray cascades either to the $5^+$ ground state or the $0^+$ isomeric state at ${\rm E_x}=228$ keV. The ground state feeding is of particular relevance for astronomy since the $^{26}$Al ground state decays into the first excited state of $^{26}$Mg with the subsequent $\gamma$-ray emission observed by the satellite telescopes. The isomeric state of $^{26}$Al decays (T$_{1/2} = 6.3$ s) exclusively to the ground state of $^{26}$Mg and does not lead to the emission of $\gamma$-rays. Therefore, a precise determination of the ground state feeding probability $f_0$ is important for the reaction rate calculation.

For the 189 and 304~keV resonances this parameter could be reinvestigated 
experimentally in a high resolution study using a high purity germanium 
detector \citep{Limata10_PRC,Strieder12_PLB}. 
A high precision determination for the low-energy resonances was 
impossible and the ground state feeding probabilities for these resonances rely 
mainly on previous literature information. The main source of information on 
the feeding probability is \citet{EN87}, which is to a large extent based on the 
experimental work published in \citet{EN88}. For resonances at 37 and 57 keV the 
feeding probability seems to be well grounded while for the 92 keV resonance there
is no experimental information from \citet{EN88}. Unfortunately, the alternative 
literature information in case of the 92~keV resonance is contradictory. 
A probability of $80\pm15$~\% was deduced from the experimental branching ratio 
determination measured in the $^{24}$Mg($^3$He,p$\gamma)^{26}$Al reaction 
\citep{CH83a,CH83b}. However, in \citet{CH86}, the same authors quote  a value 
of 61~\%, while 
the compilation of \citet{EN87} gives 85~\%. 
The origin of this large discrepancy is unknown, but may be attributed to different 
assumptions on the secondary branching ratios. Recent measurements 
\citep{Strieder12_PLB} suggested a stronger 
feeding of $^{26}$Al states that predominately decay to the isomeric state 
reducing the ground state fraction. 
Therefore, a ground state feeding probability of $60^{+20}_{-10}$~\%, as reccommended by 
\citet{Strieder12_PLB}, has been used in the present work for the 92 keV resonance. 
In general, the small uncertainty, e.g. 1~\%, quoted in \citet{EN87} seems 
questionable due to the disagreement for certain resonances and a larger uncertainty 
has been assigned to these values (see Table \ref{omegagamma_comparison}).

\subsection{Electron Screening in laboratory studies}

In astrophysical environments nuclear reactions usually take place at energies far below the Coulomb barrier where the probability for the incoming particle to overcome the repulsive force of the interacting partner decreases steeply with decreasing energy \citep{Rolfs88}. In laboratory studies the target nuclei are in most cases in the form of atoms or molecules while projectiles are usually in the form of positively charged ions. The atomic (or molecular) electron clouds surrounding the reacting nuclei act as a screening potential reducing the Coulomb barrier effectively seen by the penetrating particles. Thus, the penetration through a shielded Coulomb barrier at a given projectile energy E is equivalent to that of bare nuclei at energy $\rm E_{eff} = E + U_e$. This so called electron screening effect \citep{Assenbaum87_ZPA} becomes very important for large nuclear charges at low energies.

In general, a resonance strength $\omega\gamma$ is proportional to the penetration probability through the Coulomb barrier, the penetrability P$_l({\rm E})$ of the orbital angular momentum $l$: $\omega\gamma\propto\Gamma_{\rm p}\propto{\rm P}_l({\rm E})$. Thus, the enhancement factor f$_{\rm es}$ of the entrance channel can be expressed as:
\begin{equation}\label{screening}
{\rm f_{es}}=\frac{\omega\gamma_{\rm screen}}{\omega\gamma_{\rm bare}}=\frac{{\rm P}_l({\rm E+U_e})}{{\rm P}_l({\rm E})}
\end{equation}
and for small $l$ the approximation $\rm f_{es}\approx \exp(\pi\eta{\rm U_e/E})$ is valid \citep{Assenbaum87_ZPA} where $\eta$ is the Sommerfeld parameter \citep{Rolfs88}.
The screening potential U$_e$ is usually calculated in the approximation that the projectile velocity is much smaller than the Bohr velocity of the electrons \citep{Shoppa93_PRC}. This approximation represents the so called adiabatic limit where the electrons remain in the lowest energy state of the combined projectile and target system with the same quantum numbers as the original system. Consequently, the screening potential is given by the difference in atomic binding energy between the original system and the single positively charged combined system.

The atomic binding energies can be found in literature, e.g. \citet{Huang76_ADNDT}. In case of $^{25}$Mg(p,$\gamma$)$^{26}$Al in the adiabatic limit a value of U$_{\rm e}=1.14$~keV was calculated leading to enhancement factors f$_{\rm es}$ quoted in Table \ref{omegagamma_comparison}. However, in most cases experimental investigations of the electron screening potential resulted in larger values compared to the adiabatic limit \citep[see e.g.][]{Strieder01_NW}. This discrepancy is still far from being solved and certainly deserves further studies. It is worth noting that alternative approaches have been discussed in the literature \citep{Liolios01_NPA,Liolios03_JPG} which lead to slightly different values for the screening potential. In order to account for this ambiguity in the theoretical calculation of the electron screening potential, we assign an uncertainty to the adiabatic limit enhancement factor equal to 30~\% of the difference between its value and unity.

Note that the electron screening effect is already sizeable for the 304~keV resonance but has been totally neglected in previous compilations, e.g. \citet{angulo,Iliadis10_NPA2}, when low-energy resonance parameters from direct studies were used.

\section{The reaction rate calculation}

The Maxwellian-averaged two-body reaction rate can be calculated from \citet{Rolfs88}:
\begin{equation}
N_A\langle\sigma v\rangle=N_A\frac{(8/\pi)^{1/2}}{\mu^{1/2}(kT)^{3/2}}\int\limits_0^\infty\sigma(E)
E e^{-E/kT}dE 
\label{sig_fin}
\end{equation}
where $N_A$ is the Avogadro number, $\mu$ the reduced mass, $k$ the Boltzmann constant, $T$ the temperature, $\sigma(E)$ the cross section at the center-of-mass energy $E$, and $v$ the relative velocity of the reactants.

For narrow resonances, the reaction cross section can be expressed in the Breit-Wigner approximation and when $N_A\langle\sigma v\rangle$ is given in cm$^3$mol$^{-1}$s$^{-1}$, this leads to
\begin{equation}
N_A\langle\sigma v\rangle= \frac{1.54\times10^{11}}{(\mu T_9)^{3/2}}\sum_{i}f_{i}\omega\gamma_{i}e^{-11.605E_{i}/T_9}
\label{BW}
\end{equation}
where the energies $E$ is in MeV, $\mu$ is in amu, $T_9$ is the temperature in GK, and $(\omega\gamma)_i$ and $f_i$ are the strength (in units of MeV) and ground state feeding probability of the $i$-th resonance, respectively.

\begin{figure}
\includegraphics[angle=0,width=\columnwidth]{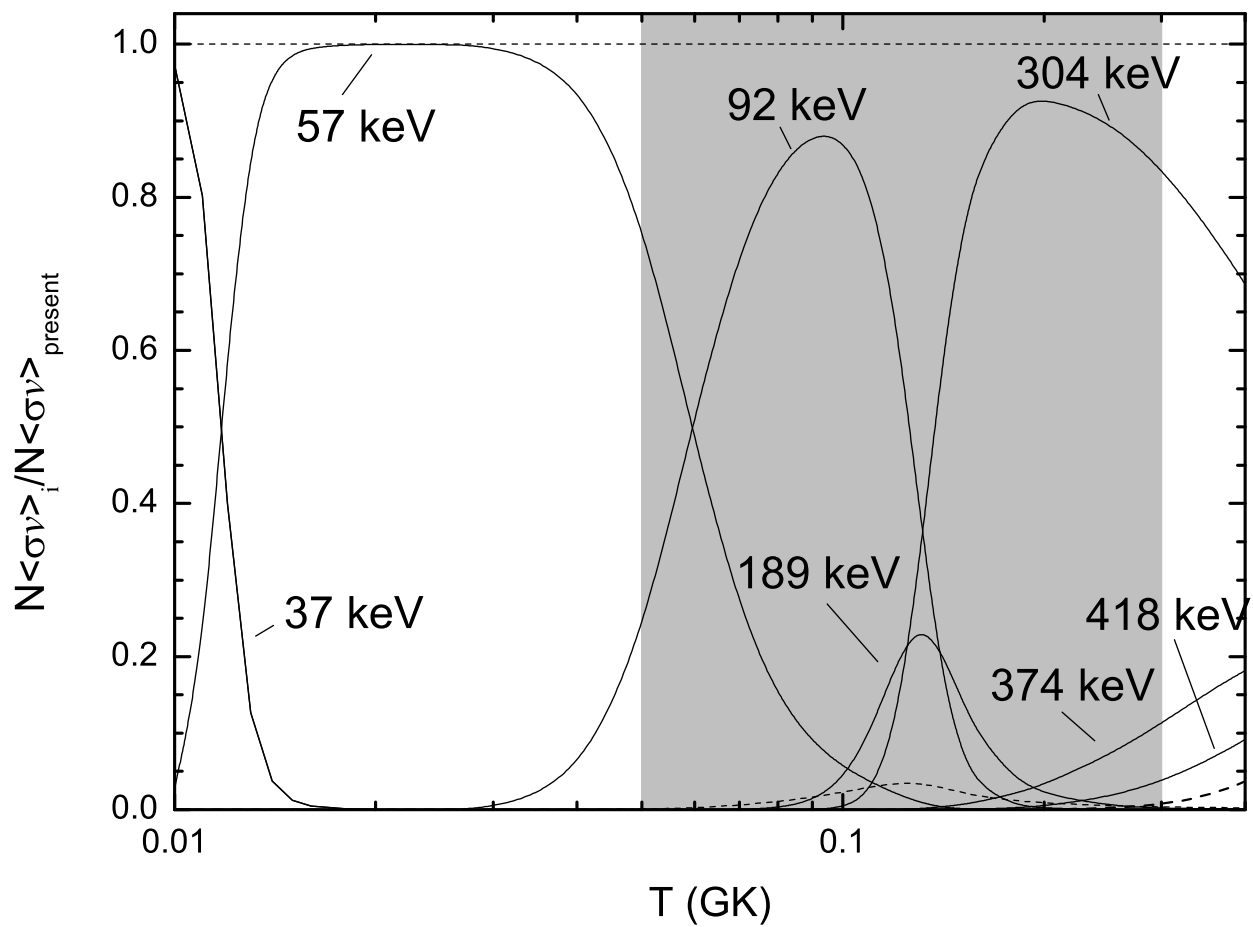}
\caption{Ratios of individual reaction rate contributions and total
recommended rate of $^{25}$Mg(p,$\gamma$)$^{26}$Al.
The dominant individual contributions are labeled while the dashed
line indicates the summed contributions of weak resonances as well as resonances above $\rm E =420$~keV.
The grey shaded area represents the temperature range for which the major revisions
have been accounted in the present work.}\label{individual}
\end{figure}

The fractional reaction rate with the contributions of individual resonances is shown in Figure \ref{individual}. The reaction rate in the temperature window between 50 and 300~MK is nearly entirely determined by the resonances measured in recent LUNA experiments \citep{Limata10_PRC,Strieder12_PLB} with a small contribution from the 57~keV resonance at the lower edge of this window. 
At larger temperatures, namely T$>$ 300 MK, the contribution of high-energy resonances becomes significant (see Figure \ref{individual}), but this temperature range is beyond the scope of the present work. 

\begin{figure}
\includegraphics[angle=0,width=\columnwidth]{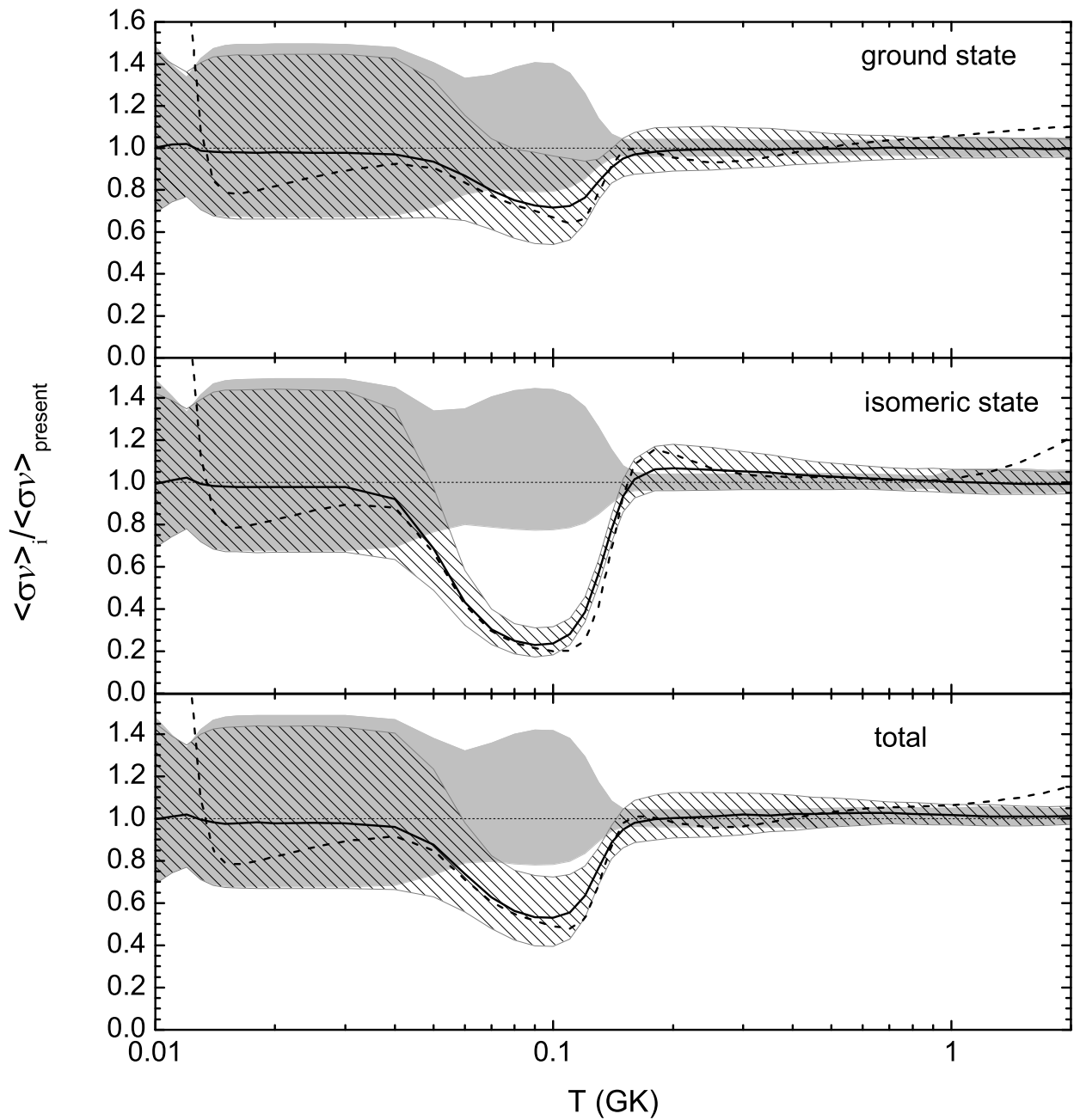}
\caption{Comparison between the present recommended reaction
rates of $^{25}$Mg(p,$\gamma$)$^{26}$Al and those reported by NACRE (dashed lines, \citet{angulo} and
\citet{Iliadis10_NPA2} (solid lines). Shaded and hatched areas represent the estimated
$1\sigma$ uncertainties 
of the present work and \citet{Iliadis10_NPA2}, respectively. Note 
that in \citet{Iliadis10_NPA2} the uncertainties on the ground state feeding factors are 
not considered.}\label{RRcomp}
\end{figure}

\begin{figure}
\includegraphics[angle=0,width=\columnwidth]{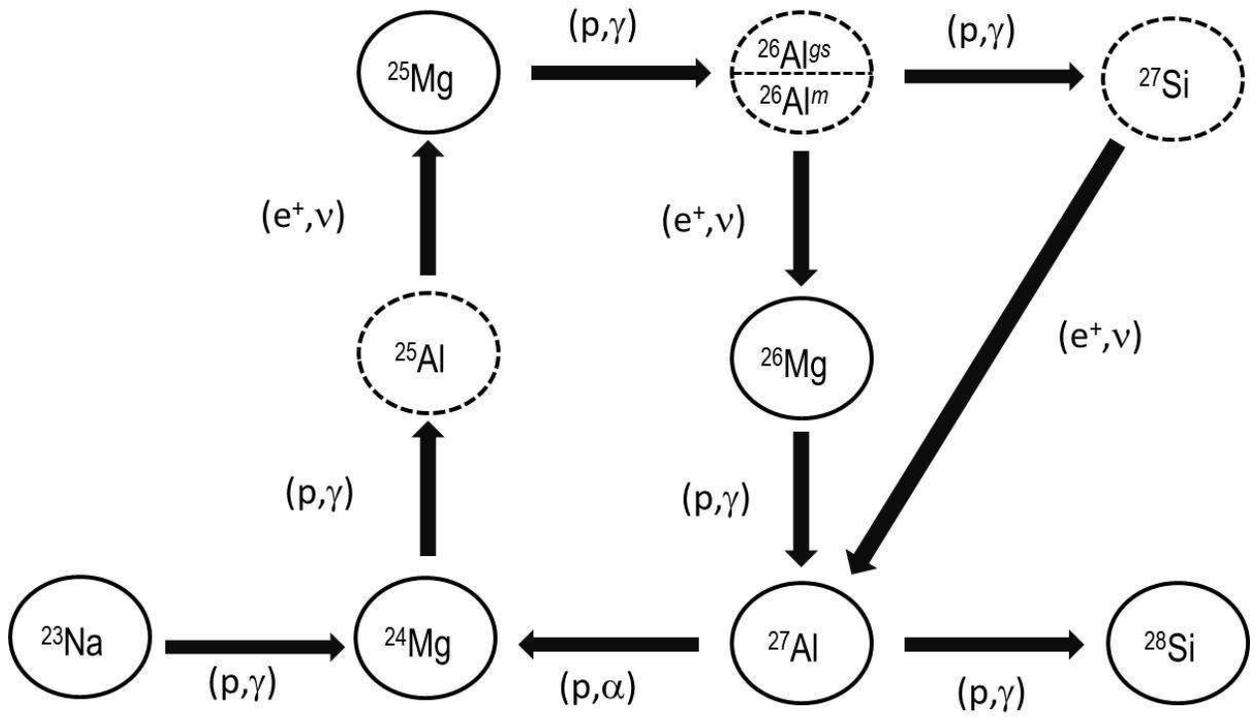}
\caption{The Mg-Al cycle: solid and dashed lines refer to stable 
and unstable isotopes, respectively.}\label{mgalcycle}
\end{figure}

The reaction rate uncertainty was investigated following the Monte Carlo approach of \citet{Iliadis10_NPA1} randomly varying the $\omega\gamma$ values entering the calculation within their experimental uncertainties. In Tables \ref{table-gs-rate} and \ref{table-m-rate} the calculated reaction rates for ground state and isomeric state are shown together with the associated lower and upper limits which are defined by the 68~\% confidence level of the obtained distribution. These new reaction rates are compared with the results of NACRE \citep{angulo} and \citet{Iliadis10_NPA2} in Figure \ref{RRcomp}.

The present reaction rates are higher than previously found because of higher $\omega\gamma$s recommended for the 92 and 189~keV resonances. In particular, the reaction rate for the isomeric state feeding increased by a factor 3-5 for temperatures between 50 and 150~MK while the ground state reaction rate is larger by 30-40~\% in the same temperature window.
The larger effect on the isomeric state reaction rate arises from the revised ground state feeding probability for the 92~keV resonance (see Sect. \ref{gsfeeding} and Table \ref{omegagamma_comparison}). 
The uncertainty at temperatures higher than $\rm T> 100$~MK is significantly reduced now due to the new accurate determination of the 304~keV resonance while at lower temperatures a sizeable uncertainty is still present. However, the parameters for the reaction rate calculation have been deeply revised in the present work and indirect data have been replaced by direct measurements when possible. Therefore, the present recommended reaction rates appear to be more robust than the results from previous work.

\section{Discussion}\label{AC}

The new rate of the $^{25}$Mg(p,$\gamma$)$^{26}$Al
is expected to produce major effects in the temperature range
$\rm 50<T<150$~MK. These conditions are typically found in the core
of massive main sequence stars as well as in the H-burning shell of RGB and AGB
stars. In this section we review three scientific cases  
related to the operation of the Mg-Al cycle in these stellar environments.
Our aim is to identify interesting problems 
of stellar evolution and nucleosynthesis whose solution requires an accurate evaluation of the  
$^{25}$Mg(p,$\gamma$)$^{26}$Al rate. 

To illustrate these scientific cases, we will make use of a bare nuclear network 
code, i.e. an appropriate set of differential equations describing the evolution 
of the abundances of all the isotopes of the Mg-Al cycle solved  under constant 
temperature and density conditions. The equations are linearized and the 
resulting set of linear equations is solved by means 
of a Newton-Rhapson algorithm. The initial abundances of Mg, Al and Si isotopes 
are taken from \citet{lodders2009} and properly scaled to the adopted metallicity.
To mimic the effect of an extended convective mixing, the H mass fraction is 
maintained constant. The adopted nuclear network is illustrated in Figure \ref{mgalcycle}.

Although a quantitative study of all the implications of the new rate would require 
the computation of appropriate stellar models, where the coupling of mixing and burning may 
be accurately accounted, a bare network calculation is adequate for most of the 
purposes of the present discussion. 
We also make use of previous results of stellar models calculations,
published in the recent literature, where the effects of a change of the reaction 
rates have been discussed in some details. 

In the following, bare network calculations obtained by means of the new rate are compared to
the ones obtained by means of the rate reccommanded by \citet{Iliadis10_NPA2}. Note that
in the quoted temperature range, the \citet{Iliadis10_NPA2} rates for the two channels of the 
$^{25}$Mg(p,$\gamma$)$^{26}$Al practically coincide with the corresponding NACRE rates.

\subsection{The $^{26}$Al in the wind of Wolf-Rayet stars}

Since the 1980s, the observations of the 1.809~MeV gamma-ray
line emitted in star forming regions of the Milky Way have raised interesting
questions about the origin of the galactic pollution of $^{26}$Al
\citep{mahoney1984,diehl95,diehl2006}.
Although it is commonly accepted that
massive stars, i.e. those ending their life as core-collapse supernovae, are
the main source of the galactic $^{26}$Al, the precise nucleosynthesis
scenario is still matter of debate.
Favoreable conditions are expected during the advanced phases of the evolution of such massive stars.
In particular, a significant contribution should come from
the pre-explosive as well as the explosive nucleosynthesis occurring in the
C- and Ne-burning shells \citep{arnett,woosley1980}. Nevertheless, 
extant theoretical models show that an additional contribution may come from Wolf-Rayet (WR) stars 
\citep{dearborn}.
In this case, the $^{26}$Al is produced within
the core of very massive main sequence stars (M$>30$ M$_\odot$),
where the temperature exceeds 50~MK.
Since the main sequence phase, these stars experience a huge mass loss. 
In such a way,
even material located on the top of the H-convective core, which is
enriched with the ashes of the Mg-Al cycle, may be ejected.
The actual contribution of the WR stars to the galactic budget
of $^{26}$Al is rather controversial. While \citet{Palacios05} find that these
stars provide between 20 to 50\% of the whole galactic $^{26}$Al,
\citet{limongi06} conclude that the cumulative yield of WRs is negligible
when compared to that from C and Ne burning shells.

\begin{figure}
\includegraphics[angle=0,width=\columnwidth]{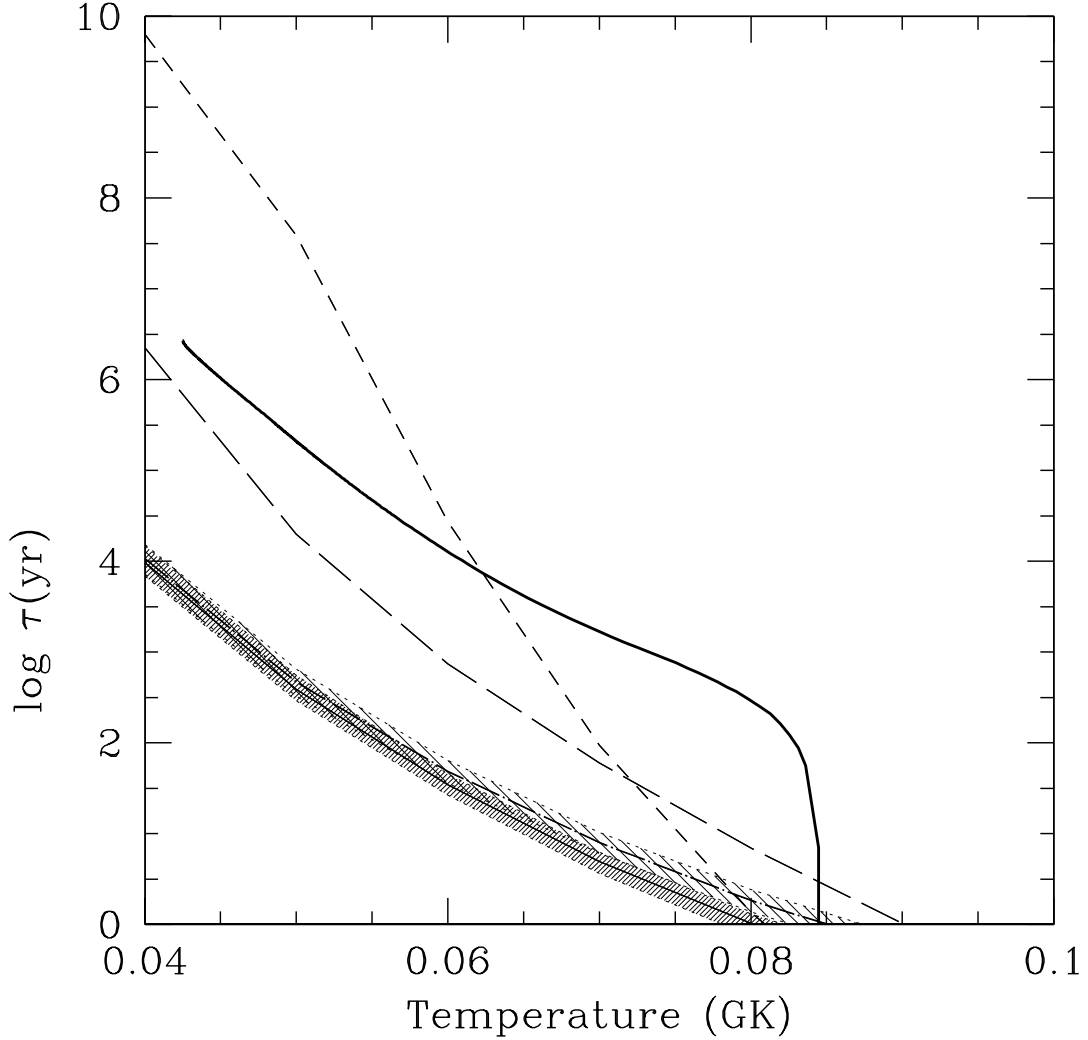}
\caption{Burning timescales of $^{25}$Mg versus temperature, as obtained by adopting the 
recommended $^{25}$Mg(p,$\gamma$)$^{26}$Al reaction rate (solid line) and the \citet{Iliadis10_NPA2} rate (dot-dashed line).
The hatched areas represent the uncertainties due to the total reaction rate.
The burning timescales of $^{24}$Mg and $^{26}$Al$^{gs}$ are also reported, dashed and long-dashed lines, respectively.
The thick-solid line represents the residual main-sequence time for a
80 M$_\odot$ models (see text for more details).}\label{tau}
\end{figure}

Figure \ref{tau} illustrates the nucleosynthesis scenario for the core-H
 burning phase of a WR precursor.
 The burning timescales of $^{24}$Mg, $^{25}$Mg, and $^{26}$Al$^{\rm gs}$
 are reported as a function of the temperature and defined as:

\begin{equation}
\tau_{i}=\frac{1}{X \rho N_A <\sigma v>_i}
\end{equation}

\noindent
where $i$ denotes for $^{24}$Mg, $^{25}$Mg, and $^{26}$Al$^{\rm gs}$ with
the corresponding reaction rates for $^{24}$Mg(p,$\gamma$)$^{25}$Al,
$^{25}$Mg(p,$\gamma$)$^{26}$Al$^{\rm tot}$, and $^{26}$Al$^{\rm gs}$(p,$\gamma$)$^{27}$Si,
respectively. We have assumed a hydrogen mass fraction $X=0.1$ and
a density of $\rho=100$~g/cm$^3$.
All the reaction rates are from the NACRE compilation except
for $^{25}$Mg(p,$\gamma$)$^{26}$Al where we have used both the present work
and the \citet{Iliadis10_NPA2} rates.
The thick solid line represents the residual time for an 80 M$_\odot$ stellar
models \citep[from]{limongi06}, i.e. the fraction of the main-sequence lifettime 
during which the central temperature is larger than a given value.

During most of the main sequence lifetime, the $^{25}$Mg(p,$\gamma$)$^{26}$Al
is the fastest process of the Mg-Al cycle.
As already found by Limongi \& Chieffi
\citep[see also][]{Iliadis11_ApJSup}, the corresponding $^{25}$Mg burning timescale
is sufficiently short to ensure that all the $^{25}$Mg available in the
convective core is converted into $^{26}$Al. Note that
only at the end of the main sequence, when the central H
is close to the complete exhaustion and the temperature is about 80~MK,
the burning rate of $^{24}$Mg becomes as short as that of $^{25}$Mg.
As a result, the burning of $^{24}$Mg provides a negligible contribution to the
$^{25}$Mg abundance in the core, and, in turn, to the $^{26}$Al production.
The $^{26}$Al$^{gs}$ accumulated in the
convective core is marginally depleted by the subsequent proton captures,
because its burning timescale is about 2 orders of magnitude
larger than that of the $^{25}$Mg.
Finally, since the $^{26}$Al$^{gs}$ lifetime
is comparable to the stellar lifetime, its radioactive decay
have to be considered.

In summary, the amount of $^{26}$Al accumulated in the convective core
of a massive star depends, essentially, on the original $^{25}$Mg content and
on the branching ratio between the two output channels of the
$^{25}$Mg(p,$\gamma$)$^{26}$Al reaction. Indeed, due to the competition between
the $^{25}$Mg(p,$\gamma$)$^{26}$Al$^{gs}$ and the $^{25}$Mg(p,$\gamma$)$^{26}$Al$^{m}$,
only a fraction of the original $^{25}$Mg is actually converted into $^{26}$Al$^{gs}$.
A comparison between the previous \citep{Iliadis10_NPA2} and the revised branching ratio
shows that at temperatures of the core H-burning, the new rates imply a substantial
increase of the competitive channel, i.e. the isomeric state production,
than previously assumed (Figure \ref{RRcomp}).
As a consequence, the $^{26}$Al$^{gs}$ production in the convective core of H-burning
massive stars is less efficient than believed so far.
Note that this finding does not necessarily imply that the contribution of WR stars
to the galactic $^{26}$Al is neglogible. 
A reliable evaluation of this contribution still resides, for example, on the 
poorly-known mass range of these stars, which is significantly affected by mass 
loss uncertainties.

\subsection{Al and Mg isotopic composition of presolar grains}

The chemical analysis of presolar grains, dust particles found in pristine meteorites with a size smaller than a few microns,
reveals a variety of isotopic compositions. These presolar grains, e.g. mainstream SiC and O-rich grains \citep[see][for a review]{hoppe-rev, clayton-rev}, represent fossil records of the parent star atmospheres and provide unique information on stellar nucleosynthesis.

Mainstream SiC grains are believed to condense in the C-rich atmospheres surrounding low-mass (M$<$3 M$_\odot$) AGB stars of different
metallicity ($0.001<Z<0.04$), which are very active nucleosynthesis sites \citep{ib1983, bgw2000, straniero06}. 
Recursive dredge-up episodes powered by thermal pulses are responsible for the C enrichment of the 
atmosphere of these giant stars and SiC grains form when the C/O ratio becomes larger than 1, the so-called C-star phase.
O-rich grains may also condense in AGB stars, but before the C-star phase is attained and C/O is still less than 1.

The $^{26}$Mg excess observed in SiC as well as O-rich grains from AGB stars is interpreted as the signature of an in-situ decay of $^{26}$Al \citep{zinner1991, nittler-nat1994} and current theoretical models predict that low-mass AGB stars may deliver
a substantial amount of $^{26}$Al. The $^{26}$Al is produced in the H-burning shell of an AGB star, accumulated in the H-exhausted region and mixed by convection powered by thermal pulses to regions of higher temperatures.
In case the maximum temperature remains below the threshold for the activation of the $^{22}$Ne($\alpha$,n)$^{25}$Mg reaction (T$<300$ MK), the $^{26}$Al survives and, later on, may be dredged up to the stellar surface. 
Contrarily, the $^{26}$Al is destroyed by neutron captures occurring at the bottom of the convective zone and only $^{26}$Al above this zone can be dredged up \citep{mowlavi2000,cristallo2009}.
Basing on full network stellar model calculations, \citet{cristallo2011} found values of $^{26}$Al/$^{27}$Al up to $5 \times 10^{-3}$, in good agreement with those measured in mainstream SiC grains and several O-rich grains. However, in some O-rich grains values larger by up to one order of magnitude have been observed.

These extreme excesses of $^{26}$Al are often explained by invoking an AGB extra-mixing which connects the bottom of the convective envelope to the hottest 
H-burning zone, where the Mg-Al cycle is at work. Note that the extra-mixing scenario provides a widely accepted explanation of 
the C and O isotopic ratios measured in the atmospheres of low-mass RGB stars \citep{boothroyd1994, charbonnel1995, denissenkov1996} and O isotopic ratios found in a 
large sample of presolar grains suggest that extra-mixing should be at work also during the AGB phase \citep{nollet2003}. 
Nevertheless, a reliable mechanism for such an extra-mixing has not been yet identified, possible candidates are rotational induced instabilities, magnetic pipes,
gravity waves and thermohaline mixing.

The AGB extra-mixing hypothesis implies that parent stars of O-rich grains with large $^{26}$Al excess
never attain the C-star stage, because otherwise one should expect also SiC grains showing similarly large values of $^{26}$Al overabundance: 
this occurrence is considered a major drawback of the proposed scenario. However, as pointed out by \citet{straniero2003}, there is a lower
limit for the mass of AGB stars with $\rm C/O>1$ and the larger the metallicity, the larger the minimum mass required, e.g. the C-star minimum mass is about 1.5 M$_\odot$ at solar metallicity, while it is only 1.3 M$_\odot$ at Z$=0.003$.
An extra-mixing in the AGB phase would increase the required minimum C-star mass, because the dredged-up carbon is partially converted into nitrogen (by the CN cycle), so that the onset of the C-star stage is delayed. Thus, a very deep extra-mixing could prevent an AGB star to become a C-star and, at the same time, would allow for the development of high excesses of $^{26}$Al.

In a recent work \cite{palmerini2011} showed that the O-rich grains with extreme $^{26}$Mg excess can be explained by AGB stellar models with particularly deep extra-mixing, provided that i) the initial mass is lower than 1.5 M$_\odot$ and ii) the $^{25}$Mg(p,$\gamma$)$^{26}$Al$^{\rm gs}$ reaction rate is enhanced by a factor of 5 with respect to the  and \citet{Iliadis10_NPA2} rates.

\begin{figure}
\includegraphics[angle=0,width=\columnwidth]{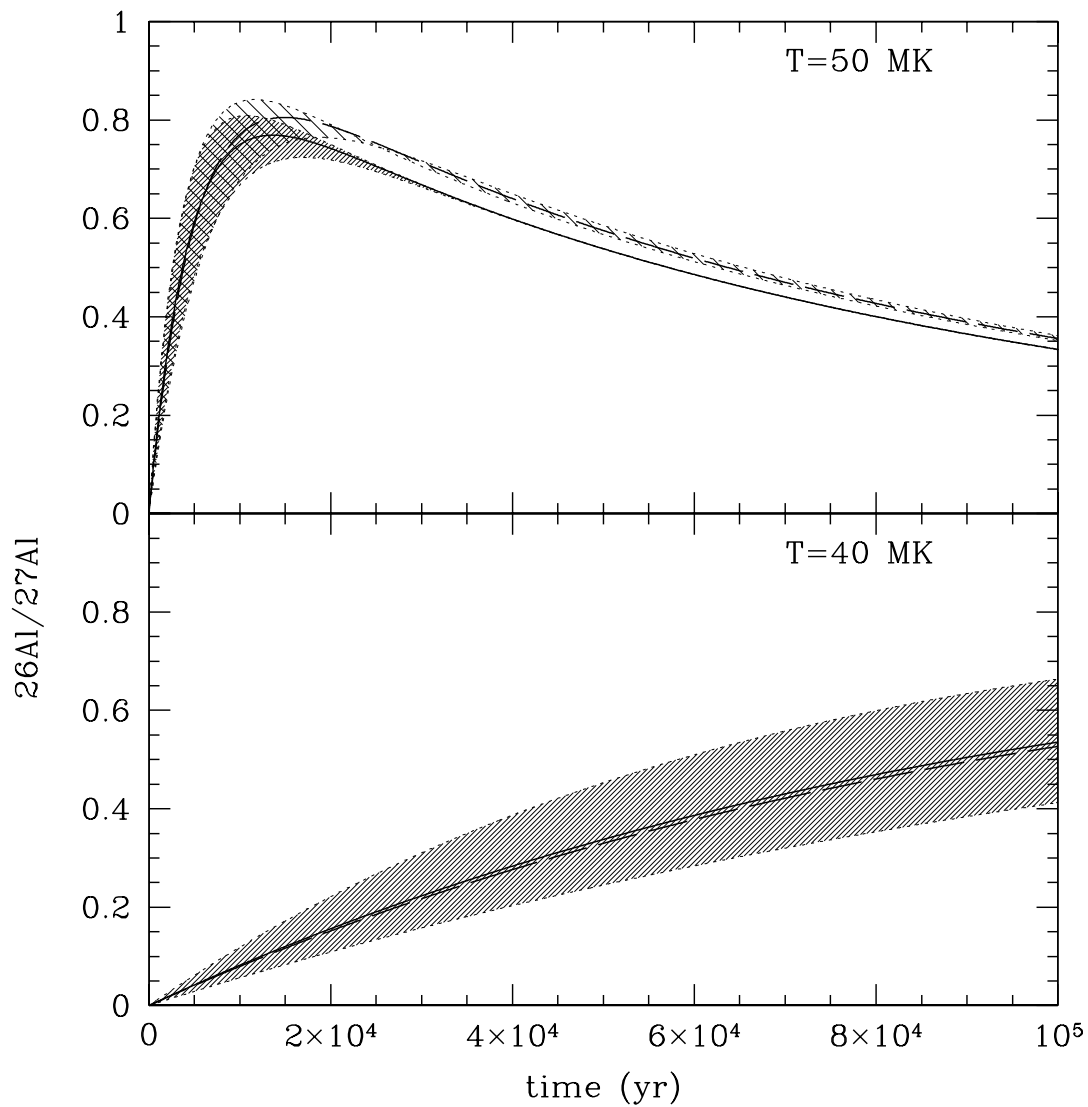}
\caption{Evolution of the aluminum isotopic ratio for material
exposed to a temperature of 40 MK (lower panel) and 50 MK (upper panel).
Solid lines represent the calculation made by means of the recommended rate of the $^{25}$Mg(p,$\gamma$)$^{26}$Al reaction, while the dashed lines have been
obtained by means of the corresponding \citet{Iliadis10_NPA2} rate.
In all cases the density is 1 g/cm$^3$,
X$=0.7$. Hatched areas represent the cumulative uncertainties due to 
both channels of the $^{25}$Mg(p,$\gamma$)$^{26}$Al reaction.}\label{alratio}
\end{figure}

The coupling between nuclear burning and mixing makes a quantitative analysis of the impact of the new rates on the
isotopic composition of O- and C-rich presolar grains difficult and would require the computation of stellar models with an extended nuclear network. 
This effort is beyond the purpose of the present work, 
but some qualitative consideration may be drawn on the basis of bare network calculations.
According to \cite{palmerini2011}, the maximum temperature attained by the extra-mixing is between 40 and 50 MK corresponding to an energy range where the $^{25}$Mg proton capture rate is dominated by the 57 keV resonance (see Figure \ref{individual}). 
In Figure \ref{alratio} we report the evolution of the $^{26}$Al/$^{27}$Al ratio for material exposed to a constant temperature of 40 MK (lower panel) and 50 MK (upper panel), respectively. In this energy range the recommended new rate for the $^{25}$Mg(p,$\gamma$)$^{26}$Al$^{\rm gs}$ is only about 10~\% larger with respect to \citet{Iliadis10_NPA2}, while the competing channel, $^{25}$Mg(p,$\gamma$)$^{26}$Al$^{\rm m}$, is about 40\% larger. As a consequence, the resulting $^{26}$Al/$^{27}$Al isotopic ratio at $\rm T=50$~MK is even lower than previously found, although the total rate is larger.
Moreover, in spite of the large uncertainty of the dominant 57 keV resonance contribution, we can definitely exclude an increase of a
factor 5 of the $^{25}$Mg(p,$\gamma$)$^{26}$Al$^{\rm gs}$ rate.

In conclusion, AGB models without extra-mixing may account for $^{26}$Al/$^{27}$Al up to 5$\times10^{-3}$, values commonly found in
mainstream SiC grains as well as in many O-rich grains from AGB stars. 
Larger values of $^{26}$Al/$^{27}$Al may be in part explained by a deep extra-mixing ($\rm T\ge 50$ MK), 
but even in the upper limit of the 
$^{25}$Mg(p,$\gamma$)$^{26}$Al$^{\rm gs}$ rate, it is unlikely that the extra-mixing scenario 
could produce aluminum isotopic ratios with $^{26}$Al/$^{27}\rm Al>10^{-2}$.

\subsection{The Mg-Al anti-correlation in Globular Clusters stars}

For many years, Globular Clusters (GCs) have been considered as simple stellar systems,
made of nearly coeval stars and formed from a chemically homogeneous
preexisting gas nebula.
Nevertheless, a growing amount of photometric and  spectroscopic observations
indicate that many GCs actually harbor multiple stellar populations
characterized by star-to-star chemical variations.
Such chemical variations include the well-known O-Na and Mg-Al anticorrelations
which are usually coupled to a nearly constant value of C+N+O.
This chemical pattern is a carachteristic signature for H-burning, 
where the Ne-Na and the Mg-Al cycles
are active. The first evidence of these "anomalies"
was found in bright red giant stars \citep{kraft, ivans1999}.
As it is well known, RGB stars have
an extended convective envelope, but the innermost unstable layer
does not reach the H-burning zone. Therefore, an extra-mixing
was initially invoked to explain the observed anticorrelations.
Nonetheless, this hypothesis is in contrast with the more recent
discovery of O-Na and the Mg-Al anticorrelations in
less evolved turn-off and sub-giant stars \citep{gratton2001, yong2003}.
These observations definitely rule out the hypothesis that the anticorrelations
are the result of an in-situ physical process and prove that they
were already present in the
gas nebula from which these stars formed about 13 Gyr ago.
Among the  proposed alternative hypothesis,
the pollution of the primordial gas by an early generation of massive AGB stars
(perhaps super-AGB) appears promising \citep{cottrell1981, dantona1983, ventura2005}.
In these massive AGB stars, the convective envelope
penetrates the regions where the H burning takes place: this phenomenon is
usually called hot bottom burning \citep{renzinivoli}.
Then, the relatively low-velocity wind of these stars   
ensures the required pollution of the intra-cluster medium.
According to this scenario, stars with low Mg and high Al
(or low O and high Na) would represent a second
generation of cluster stars, formed after the intermediate
mass stars of the first generation passed through the AGB phase and  
polluted the intra-cluster gas with ashes of H-burning.

However, the attempts made so far to simultaneously reproduce the observed
O-Na and Mg-Al anticorrelations have produced controversial results \citep{fenner2004, ventura2005}.
Recently, \citet{ventura2011}
showed that an increase of the $^{25}$Mg(p,$\gamma$)$^{26}$Al reaction rate
by a factor of 2,
coupled to a more sophisticated treatment of the convective energy transport
under super-adiabatic conditions, may reduce the discrepancy between the
theoretical expectations and the observed cluster abundances of Mg and Al.

\begin{figure}
\includegraphics[angle=0,width=\columnwidth]{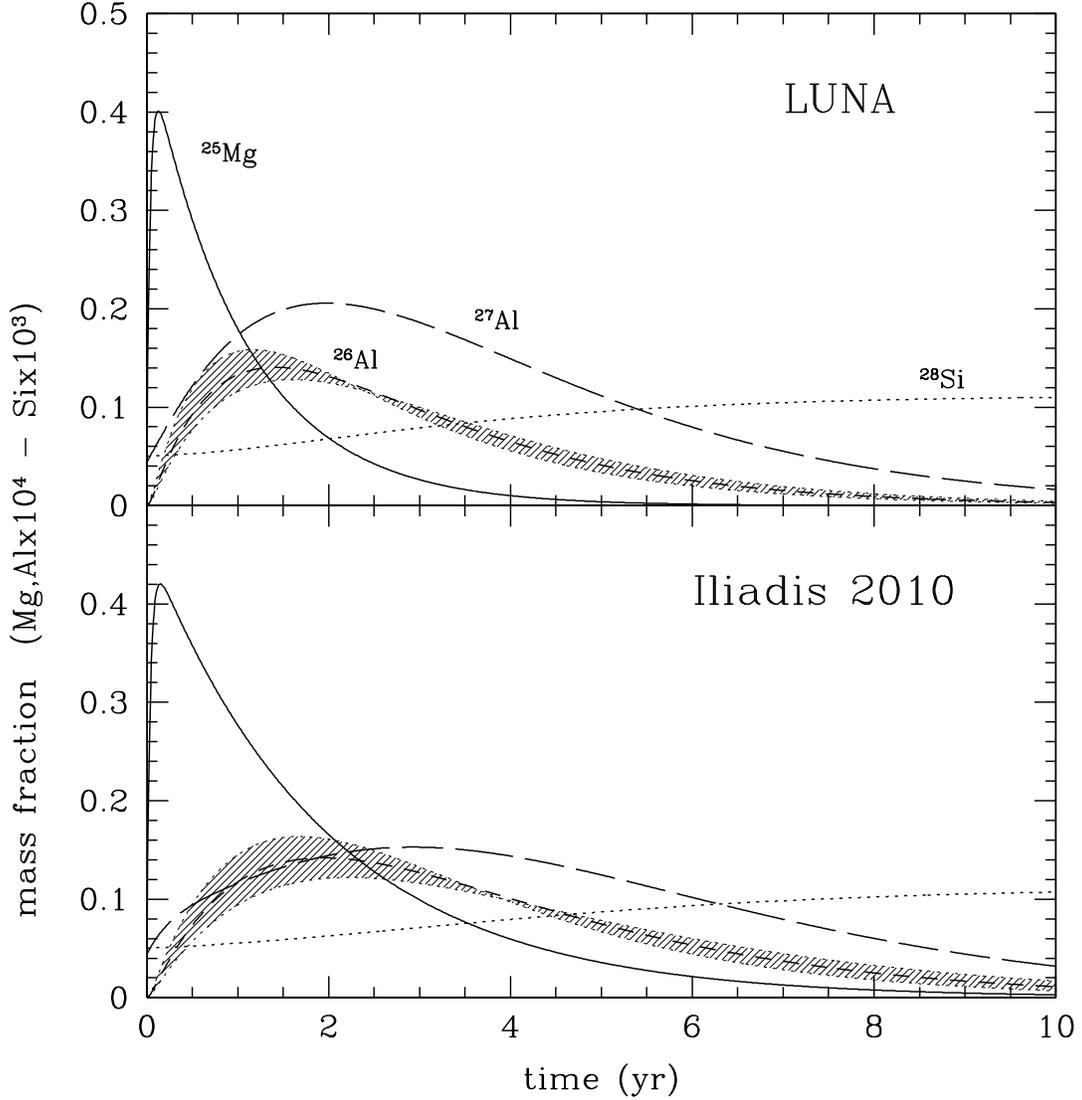}
\caption{Evolution of Mg and Al isotopes. Temperature and density are mantained constant, namely: T=100~MK and $\rho=10$ g/cm$^3$, respectively.
At t=0, the composition is scaled solar and Z=0.001. The H mass fraction is X=0.6. The various lines represent the following isotopes: $^{25}$Mg (solid), $^{26}$Al (dashed), $^{27}$Al (long dashed) and $^{28}$Si (dotted).
The calculation shown in the upper panel has been obtained by using the
recommended rates of the $^{25}$Mg(p,$\gamma$)$^{26}$Al reactions. The hatched area represents
the cumulative uncertainty on the $^{26}$Al$^{gs}$ abundance due to both 
channels of the $^{25}$Mg(p,$\gamma$)$^{26}$Al. For comparison, 
the results obtained by means of the \citet{Iliadis10_NPA2} rates are shown in the lower 
panel.}\label{100k}
\end{figure}

\begin{figure}
\includegraphics[angle=0,width=\columnwidth]{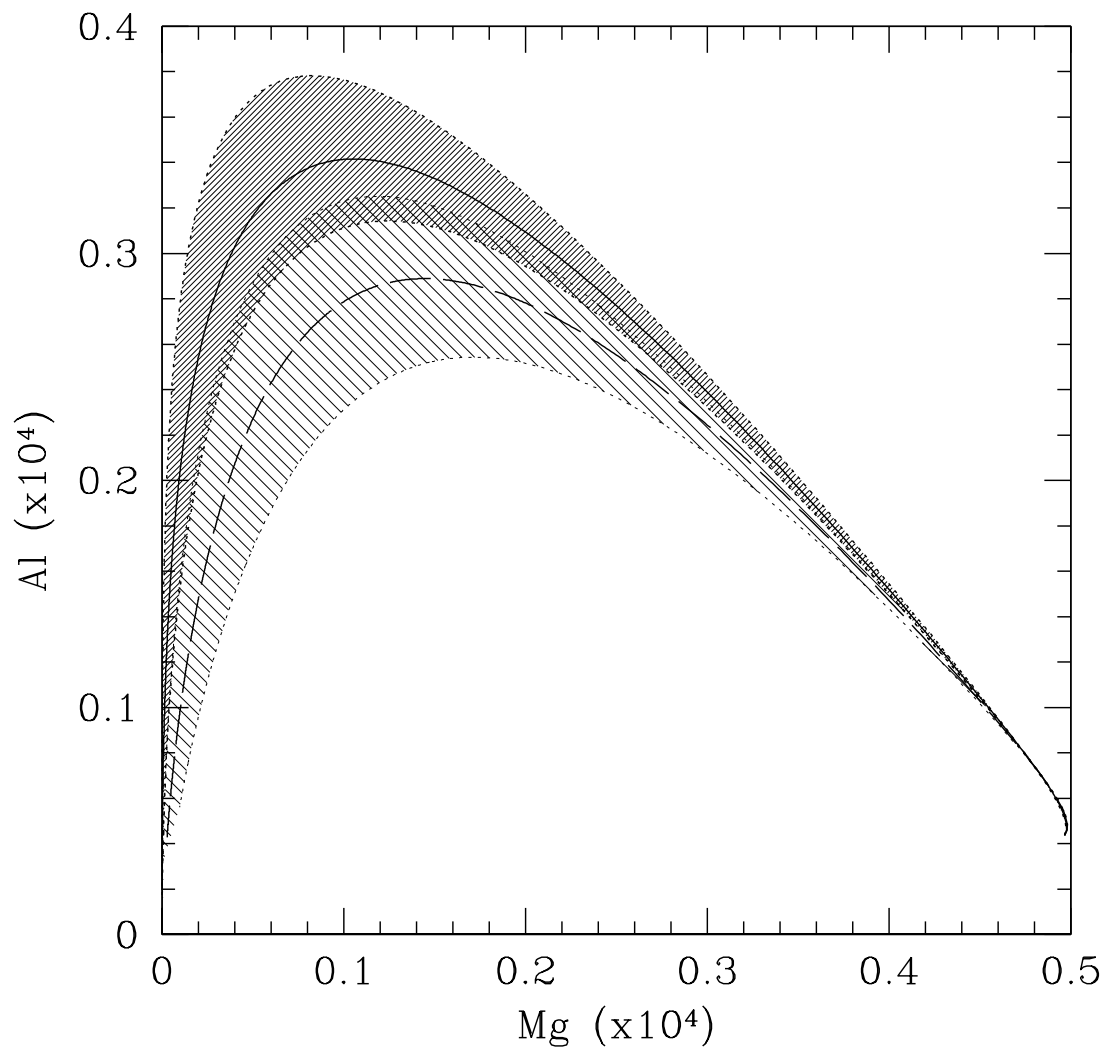}
\caption{Al abundance versus Mg abundance for the
same case shown in figure \ref{100k}.
The solid and the dashed lines refer to the calculations made by means of
the new (recommended) and the \citet{Iliadis10_NPA2} rates of the
$^{25}$Mg(p,$\gamma$)$^{26}$Al reactions, respectively.
The hatched areas represent the uncertainties due to the 
total reaction rate.}\label{anticor}
\end{figure}

In order to illustrate the influence of the newly recommended reaction rates
on the Mg-Al cycle operating in the H-burning shell of a massive AGB star,
we have performed some bare network calculations. 
Values for the metallicity, the  H mass fraction, the temperature and 
the density representative of the innermost layers of the
convective envelope of a massive AGB star have been selected, namely:
Z$=0.001$, X$=0.6$, T$=$100~MK and $\rho=10$ g/cm$^3$. 
The result is shown in Figure \ref{100k}, where
the upper panel refers to the calculation obtained by means of the new reaction
rates for the $^{25}$Mg(p,$\gamma$)$^{26}$Al, while the bottom panel corresponds
 to the calculations obtained by adopting the \citet{Iliadis10_NPA2} rates.
The NACRE compilation has been used for all the other reactions of the Mg-Al cycle,
while for the $^{26}$Al$^{gs}$ decay rate we have assumed the terrestrial
value ($\lambda=2.97\times 10^{-14}$ s$^{-1}$).
The increase of the new total $^{25}$Mg(p,$\gamma$)$^{26}$Al reaction rate
by a factor 2 with respect to \citet{Iliadis10_NPA2} is indeed very close to the value found
by \citet{ventura2011} and support the massive
AGB self-pollution scenario.
It should be noted that the largest variations in the evolution of the Mg
and Al isotopic abundances are caused by the larger $^{25}$Mg(p,$\gamma$)$^{26}$Al$^m$ rate.
This variation favors a prompt destruction of
$^{25}$Mg and a fast increase of the $^{27}$Al production.
In Figure \ref{anticor}, the Al (elemental) abundance is compared to
the corresponding Mg abundance, for the T$=100$ MK calculations.
The new rate implies a steeper anticorrelation and 
a significant increase of the maximum Al abundance. Note the similarity of this figure with 
figure 4 of \citet{ventura2011} based on the result of AGB stellar models 
obtained under different assumptions for the $^{25}$Mg(p,$\gamma$)$^{26}$Al rate.

\section{Summary and Conclusions}

The $^{25}$Mg(p,$\gamma$)$^{26}$Al reaction rate has been revised
on the basis of new measurements of the key resonances
at E=92, 189 and 304 keV.
Particular efforts have been devoted to review all experimental
parameters, e.g. resonance strengths, ground state branching ratio fractions,
and electron screening, in order to reduce the systematic uncertainty
of this reaction rate in the temperature range present in
stellar H-burning zones. 
Note that in previous works the input parameter uncertainties were partly underestimated, e.g. present uncertainties on ground state branching ratio and electron screening were not considered.

We have found a significant variation of the rate
for temperature $50<$T$<150$ MK with respect to previous studies.
The revised total reaction rate is about a factor of 2 larger
than suggested
by NACRE and \citet{Iliadis10_NPA2}, while the production rate
of the isomeric state, which decays
almost instantly into $^{26}$Mg, is up to a factor of 5 larger.
As a result, the expected production of $^{26}$Al$^{gs}$ in stellar
H-burning zones is lower than previously estimated.
This implies, in particular, a reduction of the estimated
contribution of WR stars to the galactic production of $^{26}$Al.
We have also investigated the possible effect on the
Mg and Al isotopic composition of presolar grains originated in AGB stars.
The most important conclusion is that the deep AGB extra-mixing, often invoked to
explain the large excess of $^{26}$Al in some O-rich grains, does not appear
a suitable solution for $^{26}$Al/$^{27}$Al$>10^{-2}$.

On the other hand, the substantial increase of the total reaction rate
makes the Globular Cluster self-pollution caused by massive AGB stars
a more reliable scenario for the reproduction of the Mg-Al
anticorrelation.

In summary, we have demonstrated that a considerable improvement of our
knowledge of the nuclear reaction rates involved in the Mg-Al cycle
allows to constrain nucleosynthesis
and stellar evolution models as well as the
interplay between nuclear
burning and mixing processes operating simultaneously in stellar interiors.
In this context, further experimental studies are required to improve the
analysis reported in section \ref{AC} and to derive more firm conclusions on the 
operation of the Mg-Al cycle in stellar interiors. 
Some input parameters still carry a significant uncertainty, e.g. the ground state branching ratio of each nuclear resonance.
Partially, as in case of the 92 keV resonance, these branching ratios are based on experiments with rather low statistics and, therefore, we recommend a reinvestigation of these parameters in a dedicated experiment.
In addition, other key reactions of the Mg-Al cycle, such as the $^{24}$Mg(p,$\gamma$)$^{25}$Al, deserve more attention.
Note that $^{24}$Mg is the most abundant isotopes among those involved in the Mg-Al cycle and  
at T$>$80 MK this reaction is faster than 
$^{25}$Mg(p,$\gamma$)$^{26}$Al, thus providing additional fuel for the Al
production. The rate tabulated by NACRE is essentially based on the experimental
result by \citet{trautrolfs}. At low energy, the cross section is dominated 
by a resonance at 214 keV. An experiment performed by of the TUNL group 
\citep{tunl1999} resulted in a 25\% higher resonance strength than recommended 
by NACRE. Note that this result has been incorporated by \citet{Iliadis10_NPA2}
in their revised reaction rate. 
However, \citet{Limata10_PRC} derive a value for the strength of the 214 KeV 
resonance that agrees with the old result by \citet{trautrolfs}. 
Further studies are required to disentangle these controversial results.
Concerning the production of $^{26}$Al$^{gs}$, a key role is played by 
the $^{26}$Al$^{gs}$(p,$\gamma$)$^{27}$Si reaction.
The only available direct measurement has been discussed
in \citet{Vogelaar1996}. Recently, the 184 keV resonance has been measured at 
TRIUMF with the Recoil Mass Separator \citep{ruiz2006}. 
An up-to-date analysis of the reaction rate has been presented by \citet{Iliadis10_NPA2}.
The difficulties of these measurements are related to the radioactivity 
of $^{26}$Al$^{gs}$. Also in this case further experimental investigations are mandatory.

\acknowledgments

The present work has been supported by INFN and in part by the EU
(ILIAS-TA RII3-CT-2004-506222), OTKA (K101328), and DFG (Ro~429/41).
We are grateful to M. Limongi for the enlightening discussions on the $^{26}$Al 
production in massive stars.
A. Di Leva, G. Imbriani, L. Piersanti and S. Cristallo aknowledge the
support of the Italian Ministry of Education, University and Research
under the FIRB2008 program.
O. Straniero, L. Piersanti and S. Cristallo
have been supported by INAF under the PRIN2010 program. A. Caciolli aknowledges 
financial support by Fondazione Cassa Di Risparmio di Padova e Rovigo.

\bibliographystyle{apj}
\bibliography{letter}

\clearpage

\begin{deluxetable}{cccccccc}
\tabletypesize{\scriptsize}
\tablecaption{The new recommended $^{25}\rm Mg(p,\gamma)^{26}Al$ resonance strengths (uncorrected for screening) and corresponding ground state fractions $f_0$. The parameters for resonances not listed here were taken from \citet{Iliadis10_NPA3}. The electron screening enhancement factor $\rm f_{es}$ was calculated according to \citet{Assenbaum87_ZPA}.\label{omegagamma_comparison}}
\tablewidth{0pt}
\tablehead{
  & \multicolumn{3}{c}{present work} &  & \multicolumn{2}{c}{\citet{Iliadis10_NPA2}} & \citet{angulo}(NACRE)\tablenotemark{a} \\
 \cline{2-4} \cline{6-7}  \\
 \colhead{E (keV)\tablenotemark{b}} & \colhead{$\omega\gamma$ (eV)} & \colhead{f$_{\rm es}$} & \colhead{$f_0$} & & \colhead{$\omega\gamma$ (eV)} & \colhead{$f_0\tablenotemark{c}$} & \colhead{$\omega\gamma$ (eV)}}
\startdata
  37.0  & $(4.5\pm1.8)\times10^{-22}$\tablenotemark{d}   &  -   &  $0.79\pm0.05$\tablenotemark{e} & & $(4.5\pm1.8)\times10^{-22}$   & 0.79 &  $(2.4^{+21.6}_{-2.4})\times10^{-21}$ \\
  57.4  & $(2.8\pm1.1)\times10^{-13}$\tablenotemark{d}   &  -   &  $0.81\pm0.05$\tablenotemark{e} & & $(2.8\pm1.1)\times10^{-13}$   & 0.81 &  $(2.82^{+1.41}_{-0.94})\times10^{-13}$ \\
  92.2  & $(2.9\pm0.6)\times10^{-10}$\tablenotemark{f}   & $1.25\pm0.08$ & $0.6^{+0.2}_{-0.1}$\tablenotemark{f} & & $(1.16\pm0.46)\times10^{-10}$  & 0.85 & $(1.16^{+1.16}_{-0.39})\times10^{-10}$ \\
  189.5 & $(9.0\pm0.6)\times10^{-7}$\tablenotemark{f}    & $1.08\pm0.03$ & $0.75\pm0.02$\tablenotemark{f} &  & $(7.2\pm1.0)\times10^{-7}$    & 0.66 & $(7.1\pm0.9)\times10^{-7}$ \\
  304.0 & $(3.08\pm0.13)\times10^{-2}$\tablenotemark{g}  & $1.04\pm0.01$ & $0.878\pm0.010$\tablenotemark{g} &  & $(3.0\pm0.4)\times10^{-2}$  & 0.87 & $(3.1\pm0.2)\times10^{-2}$ \\
\enddata
\tablenotetext{a}{the numerical values used for the ground state feeding probability are not provided}
\tablenotetext{b}{from \citet{EN87}, the uncertainty is less than 0.2~keV in all cases}
\tablenotetext{c}{from \citet{EN87}}
\tablenotetext{d}{from \citet{Iliadis10_NPA3}}
\tablenotetext{e}{from \citet{EN87} where a larger uncertainty than originally quoted was assumed}
\tablenotetext{f}{from \citet{Strieder12_PLB}}
\tablenotetext{g}{from \citet{Limata10_PRC}}
\end{deluxetable}

\clearpage

\begin{deluxetable}{cccc}
\tablecolumns{4}
\tablewidth{0pc}
\tablecaption{Reaction rate for $^{25}\rm Mg(p,\gamma)^{26}Al^{gs}$ (cm$^3$ mol$^{-1}$ s$^{-1}$).\label{table-gs-rate}}
\tablehead{\colhead{T (GK)} &  \colhead{lower limit}  &  \colhead{recommended value} & \colhead{upper limit}}
\startdata
0.010	&   $8.23\times10^{-33}$  &   $1.22\times10^{-32}$  &   $1.81\times10^{-32}$ \\
0.011	&   $4.58\times10^{-31}$  &   $6.37\times10^{-31}$  &   $8.89\times10^{-31}$ \\
0.012	&   $2.21\times10^{-29}$  &   $2.95\times10^{-29}$  &   $3.95\times10^{-29}$ \\
0.013	&   $9.24\times10^{-28}$  &   $1.32\times10^{-27}$  &   $1.88\times10^{-27}$ \\
0.014	&   $2.87\times10^{-26}$  &   $4.21\times10^{-26}$  &   $6.20\times10^{-26}$ \\
0.015	&   $6.00\times10^{-25}$  &   $8.90\times10^{-25}$  &   $1.32\times10^{-24}$ \\
0.016	&   $8.71\times10^{-24}$  &   $1.30\times10^{-23}$  &   $1.93\times10^{-23}$ \\
0.018	&   $7.50\times10^{-22}$  &   $1.12\times10^{-21}$  &   $1.67\times10^{-21}$ \\
0.020	&   $2.61\times10^{-20}$  &   $3.89\times10^{-20}$  &   $5.82\times10^{-20}$ \\
0.025	&   $1.49\times10^{-17}$  &   $2.21\times10^{-17}$  &   $3.31\times10^{-17}$ \\
0.030	&   $9.70\times10^{-16}$  &   $1.45\times10^{-15}$  &   $2.16\times10^{-15}$ \\
0.040	&   $1.71\times10^{-13}$  &   $2.52\times10^{-13}$  &   $3.72\times10^{-13}$ \\
0.050	&   $4.28\times10^{-12}$  &   $5.96\times10^{-12}$  &   $8.39\times10^{-12}$ \\
0.060	&   $4.86\times10^{-11}$  &   $6.25\times10^{-11}$  &   $8.32\times10^{-11}$ \\
0.070	&   $3.34\times10^{-10}$  &   $4.17\times10^{-10}$  &   $5.62\times10^{-10}$ \\
0.080	&   $1.54\times10^{-9}$  &   $1.93\times10^{-9}$  &   $2.68\times10^{-9}$ \\
0.090	&   $5.28\times10^{-9}$  &   $6.68\times10^{-9}$  &   $9.39\times10^{-9}$ \\
0.100	&   $1.48\times10^{-8}$  &   $1.87\times10^{-8}$  &   $2.62\times10^{-8}$ \\
0.110	&   $3.82\times10^{-8}$  &   $4.70\times10^{-8}$  &   $6.39\times10^{-8}$ \\
0.120	&   $1.05\times10^{-7}$  &   $1.23\times10^{-7}$  &   $1.55\times10^{-7}$ \\
0.130	&   $3.53\times10^{-7}$  &   $3.87\times10^{-7}$  &   $4.41\times10^{-7}$ \\
0.140	&   $1.38\times10^{-6}$  &   $1.45\times10^{-6}$  &   $1.55\times10^{-6}$ \\
0.150	&   $5.41\times10^{-6}$  &   $5.63\times10^{-6}$  &   $5.87\times10^{-6}$ \\
0.160	&   $1.93\times10^{-5}$  &   $2.01\times10^{-5}$  &   $2.09\times10^{-5}$ \\
0.180	&   $1.75\times10^{-4}$  &   $1.82\times10^{-4}$  &   $1.90\times10^{-4}$ \\
0.200	&   $1.05\times10^{-3}$  &   $1.09\times10^{-3}$  &   $1.14\times10^{-3}$ \\
0.250	&   $2.64\times10^{-2}$  &   $2.75\times10^{-2}$  &   $2.86\times10^{-2}$ \\
0.300	&   $2.26\times10^{-1}$  &   $2.35\times10^{-1}$  &   $2.44\times10^{-1}$ \\
0.350	&   $1.05\times10^{0}$  &   $1.09\times10^{0}$  &   $1.13\times10^{0}$ \\
0.400	&   $3.34\times10^{0}$  &   $3.46\times10^{0}$  &   $3.59\times10^{0}$ \\
0.450   &   $8.29\times10^{0}$  &   $8.59\times10^{0}$  &   $8.89\times10^{0}$ \\
0.500   &   $1.73\times10^{1}$  &   $1.79\times10^{1}$  &   $1.85\times10^{1}$ \\
0.600   &   $5.28\times10^{1}$  &   $5.48\times10^{1}$  &   $5.67\times10^{1}$ \\
0.700   &   $1.20\times10^{2}$  &   $1.24\times10^{2}$  &   $1.28\times10^{2}$ \\
0.800   &   $2.23\times10^{2}$  &   $2.31\times10^{2}$  &   $2.40\times10^{2}$ \\
0.900   &   $3.66\times10^{2}$  &   $3.79\times10^{2}$  &   $3.93\times10^{2}$ \\
1.000   &   $5.40\times10^{2}$  &   $5.66\times10^{2}$  &   $5.95\times10^{2}$ \\
1.250   &   $1.14\times10^{3}$  &   $1.19\times10^{3}$  &   $1.25\times10^{3}$ \\
1.500   &   $1.89\times10^{3}$  &   $1.98\times10^{3}$  &   $2.07\times10^{3}$ \\
1.750   &   $2.78\times10^{3}$  &   $2.91\times10^{3}$  &   $3.05\times10^{3}$ \\
2.000   &   $3.77\times10^{3}$  &   $3.92\times10^{3}$  &   $4.10\times10^{3}$ \\
\enddata
\end{deluxetable}

\clearpage

\begin{deluxetable}{c c c c}
\tablecolumns{4}
\tablewidth{0pc}
\tablecaption{Reaction rate for $^{25}\rm Mg(p,\gamma)^{26}Al^{m}$ (cm$^3$ mol$^{-1}$ s$^{-1}$).\label{table-m-rate}}
\tablehead{\colhead{T (GK)} &  \colhead{lower limit}  &  \colhead{recommended value} & \colhead{upper limit}}
\startdata
0.010	&   $2.31\times10^{-33}$  &   $3.43\times10^{-33}$  &   $5.09\times10^{-33}$ \\
0.011	&   $1.22\times10^{-31}$  &   $1.72\times10^{-31}$  &   $2.42\times10^{-31}$ \\
0.012	&   $5.58\times10^{-30}$  &   $7.42\times10^{-30}$  &   $9.89\times10^{-30}$ \\
0.013	&   $2.23\times10^{-28}$  &   $3.14\times10^{-28}$  &   $4.45\times10^{-28}$ \\
0.014	&   $6.79\times10^{-27}$  &   $9.89\times10^{-27}$  &   $1.45\times10^{-26}$ \\
0.015	&   $1.41\times10^{-25}$  &   $2.08\times10^{-25}$  &   $3.09\times10^{-25}$ \\
0.016	&   $2.05\times10^{-24}$  &   $3.03\times10^{-24}$  &   $4.50\times10^{-24}$ \\
0.018	&   $1.76\times10^{-22}$  &   $2.61\times10^{-22}$  &   $3.88\times10^{-22}$ \\
0.020	&   $6.14\times10^{-21}$  &   $9.09\times10^{-21}$  &   $1.35\times10^{-20}$ \\
0.025	&   $3.49\times10^{-18}$  &   $5.16\times10^{-18}$  &   $7.70\times10^{-18}$ \\
0.030	&   $2.28\times10^{-16}$  &   $3.38\times10^{-16}$  &   $5.03\times10^{-16}$ \\
0.040	&   $4.28\times10^{-14}$  &   $6.17\times10^{-14}$  &   $8.93\times10^{-14}$ \\
0.050	&   $1.41\times10^{-12}$  &   $1.84\times10^{-12}$  &   $2.46\times10^{-12}$ \\
0.060	&   $2.12\times10^{-11}$  &   $2.66\times10^{-11}$  &   $3.58\times10^{-11}$ \\
0.070	&   $1.72\times10^{-10}$  &   $2.19\times10^{-10}$  &   $3.07\times10^{-10}$ \\
0.080	&   $8.73\times10^{-10}$  &   $1.12\times10^{-9}$  &   $1.61\times10^{-9}$ \\
0.090	&   $3.14\times10^{-9}$  &   $4.06\times10^{-9}$  &   $5.87\times10^{-9}$ \\
0.100	&   $8.88\times10^{-9}$  &   $1.15\times10^{-8}$  &   $1.65\times10^{-8}$ \\
0.110	&   $2.16\times10^{-8}$  &   $2.76\times10^{-8}$  &   $3.90\times10^{-8}$ \\
0.120	&   $4.98\times10^{-8}$  &   $6.17\times10^{-8}$  &   $8.37\times10^{-8}$ \\
0.130	&   $1.21\times10^{-7}$  &   $1.43\times10^{-7}$  &   $1.80\times10^{-7}$ \\
0.140	&   $3.40\times10^{-7}$  &   $3.77\times10^{-7}$  &   $4.35\times10^{-7}$ \\
0.150	&   $1.06\times10^{-6}$  &   $1.12\times10^{-6}$  &   $1.21\times10^{-6}$ \\
0.160	&   $3.34\times10^{-6}$  &   $3.48\times10^{-6}$  &   $3.64\times10^{-6}$ \\
0.180	&   $2.74\times10^{-5}$  &   $2.85\times10^{-5}$  &   $2.95\times10^{-5}$ \\
0.200	&   $1.62\times10^{-4}$  &   $1.68\times10^{-4}$  &   $1.74\times10^{-4}$ \\
0.250	&   $4.26\times10^{-3}$  &   $4.42\times10^{-3}$  &   $4.59\times10^{-3}$ \\
0.300	&   $3.89\times10^{-2}$  &   $4.04\times10^{-2}$  &   $4.20\times10^{-2}$ \\
0.350	&   $1.93\times10^{-1}$  &   $2.01\times10^{-1}$  &   $2.09\times10^{-1}$ \\
0.400	&   $6.53\times10^{-1}$  &   $6.81\times10^{-1}$  &   $7.10\times10^{-1}$ \\
0.450   &   $1.72\times10^{0}$  &   $1.79\times10^{0}$  &   $1.87\times10^{0}$ \\
0.500   &   $3.79\times10^{0}$  &   $3.95\times10^{0}$  &   $4.12\times10^{0}$ \\
0.600   &   $1.29\times10^{1}$  &   $1.34\times10^{1}$  &   $1.40\times10^{1}$ \\
0.700   &   $3.21\times10^{1}$  &   $3.33\times10^{1}$  &   $3.45\times10^{1}$ \\
0.800   &   $6.52\times10^{1}$  &   $6.74\times10^{1}$  &   $6.96\times10^{1}$ \\
0.900   &   $1.15\times10^{2}$  &   $1.18\times10^{2}$  &   $1.22\times10^{2}$ \\
1.000   &   $1.77\times10^{2}$  &   $1.87\times10^{2}$  &   $1.98\times10^{2}$ \\
1.250   &   $4.17\times10^{2}$  &   $4.40\times10^{2}$  &   $4.68\times10^{2}$ \\
1.500   &   $7.55\times10^{2}$  &   $7.94\times10^{2}$  &   $8.46\times10^{2}$ \\
1.750   &   $1.17\times10^{3}$  &   $1.23\times10^{3}$  &   $1.30\times10^{3}$ \\
2.000   &   $1.64\times10^{3}$  &   $1.73\times10^{3}$  &   $1.83\times10^{3}$ \\
\enddata
\end{deluxetable}

\clearpage

\clearpage

\clearpage

\clearpage

\clearpage

\clearpage

\clearpage

\end{document}